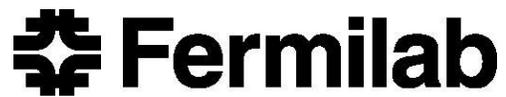

# COMMISSIONING REPORT OF THE MUCOOL 5 TESLA SOLENOID COUPLED WITH HELIUM REFRIGERATOR

MARCH - MAY 2010

ENGINEERING NOTE
FERMILAB-TM-2462-AD

Michael Geynisman
Fermi National Accelerator Laboratory
May 17, 2010

**Table of Contents**

# I Acknowledgements

# II Commissioning Schedule

This report describes results of the commissioning of the MuCool refrigeration system coupled with superconducting 5T solenoid. The commissioning was done from March 4 through April 1, 2010.

**Table 1**

| # | Dates | Activities |
|---|-------|------------|
| 1 | March 4-12 | Purification of the helium system, including compressor, refrigerator, transfer line and valve box, solenoid and cooldown piping |
| 2 | March 18-19 | Cooldown of the refrigerator and fill of the 2ph separator |
| 3 | March 19-21 | Cooldown and fill of the transfer line and solenoid to 30% liquid helium level |
| 4 | March 21-24 | Tuning of the system and fill of the solenoid to 70-75% liquid helium level |
| 5 | March 25-27 | Power test of the solenoid and quench response |
| 6 | March 27-29 | Steady state operations in JT mode with wet engine off |
| 7 | March 29-31 | Isolating fill valve to the solenoid and monitoring boil-off rate |
| 8 | March 31-April 1 | Warm up of the system and shutting off operations |
| 8 | May 3-10 | Purification and start the system for long-term magnet operations |

There were no safety incidents during the commissioning run, except a helium leak and momentary oxygen deficiency registered in the hall during the solenoid quench and venting helium to the outside via vent manifold. This incident is described later in the report. All process alarms and interlocks, as well as ODH and fire alarms, were active and performed as designed. No significant cryogenic or vacuum leaks were registered. The summary can be found in the conclusions section.

# III General Information and Summary of previous data for Solenoid and Helium Refrigerator

The MuCool Experiment [1,2,3] is designed to take data with 805 and 201 MHz cavities in the MuCool Test Area. The system uses RF power sources from the Fermilab Linac. MuCool Experiment has been studying the dependence of RF limits on frequency, cavity material, high magnetic fields, gas pressure, coatings, etc. with the general aim of understanding the basic mechanisms involved. In order to test high gradient RF of up to 40 MV/meter cavities in a magnetic field of up to 5 T, a superconducting solenoid magnet was designed and built by Wang NMR. The solenoid is enclosed in the helium cooled cryostat shown on Fig. 1.

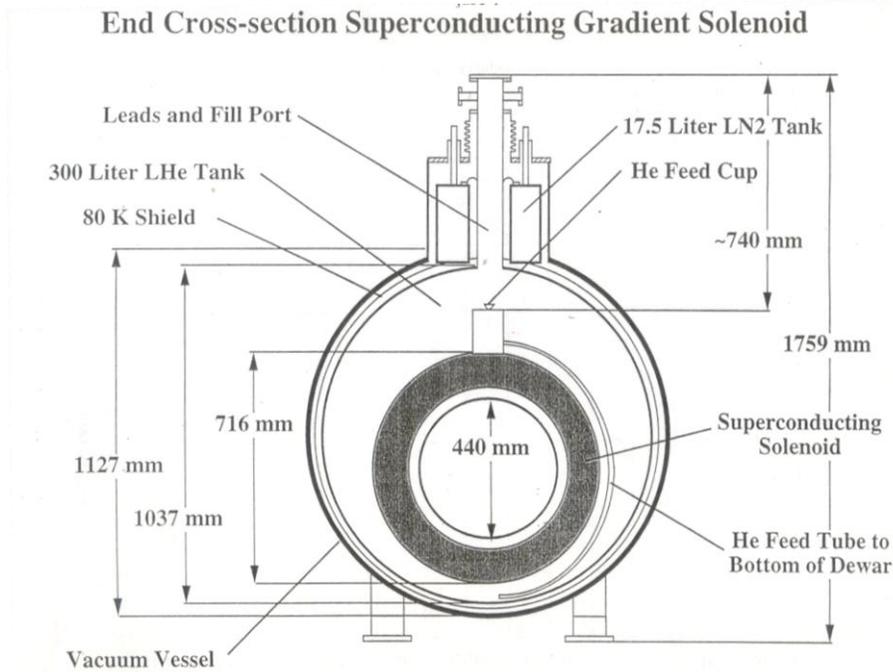

**Figure 1 MTA Hall 5T Solenoid**

Since 2005 the solenoid and a combination of RF cavities have installed in the experimental hall (see Fig. 2). Until recently the solenoid has been cooled down with combination of liquid nitrogen and liquid helium from portable dewars. A standard cooldown procedure would use three to four 180L liquid nitrogen tanks to cool the coil down to 90-100K utilizing large latent heat of nitrogen. Then the nitrogen supply would have to be pulled from the solenoid's fill connection and replaced with the supply from the 500L liquid helium dewars. Then helium would be used to purge nitrogen from the system and cooldown the coil to 4K (see Fig.3). These operations were labor intensive; required experts to insert fill bayonet properly to match the helium collection cap receptacle 29 inches below the cryostat top flange (see Fig.1); stressed components of the fill connections; required careful purging nitrogen with helium; and finally wasted cooldown helium to atmosphere. Consequently, all helium required to maintain solenoid at operating temperature was lost to atmosphere.

It is estimated that up to 1,500 liters of helium would be required for cooldown, plus up to 100 liters a day to maintain boil-off rate. Therefore a month-long operation would waist 4,500 liters of helium. In 2006 the experiment ordered 35,500 liters of helium. Most importantly, in order to maintain solenoid cold, two technicians were assigned to monitor parameters, order cryogens and maintain levels.

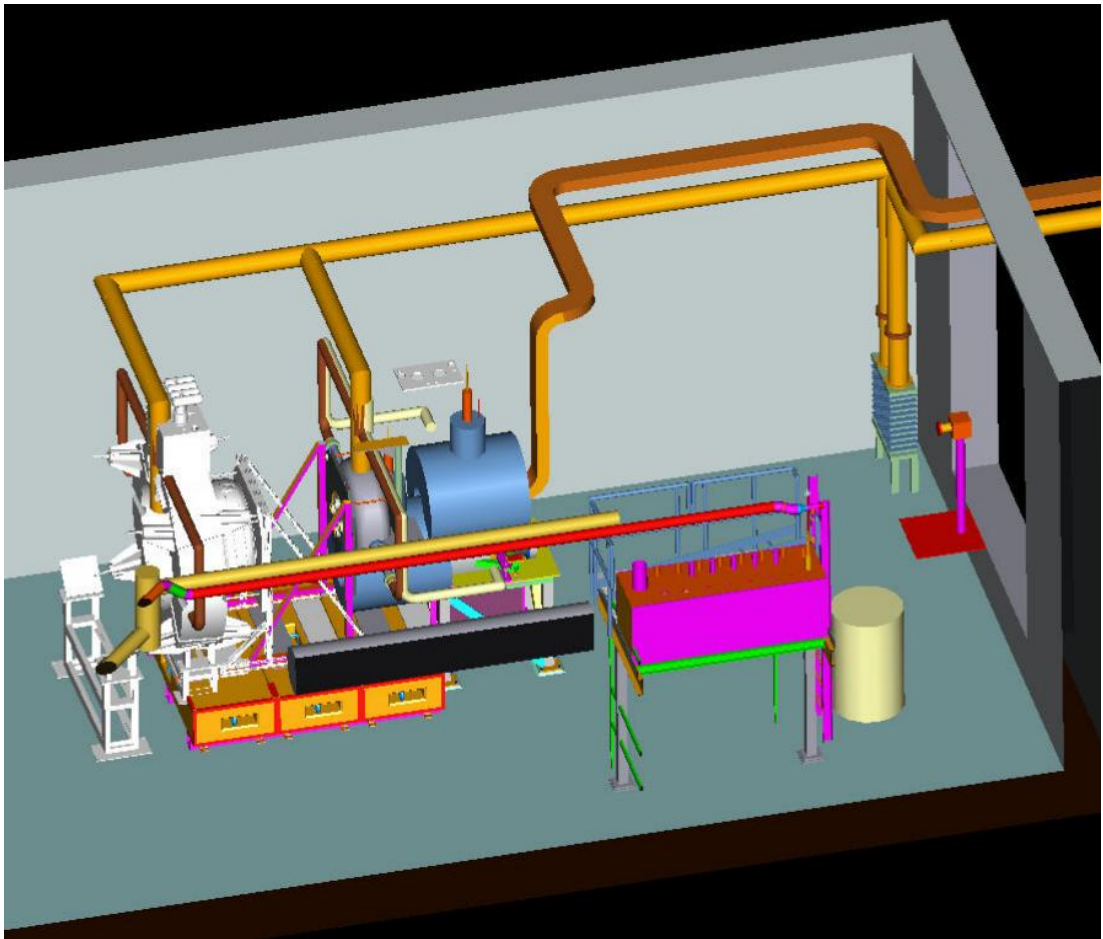

**Figure 2 Present (2010) configuration of the 5T solenoid**

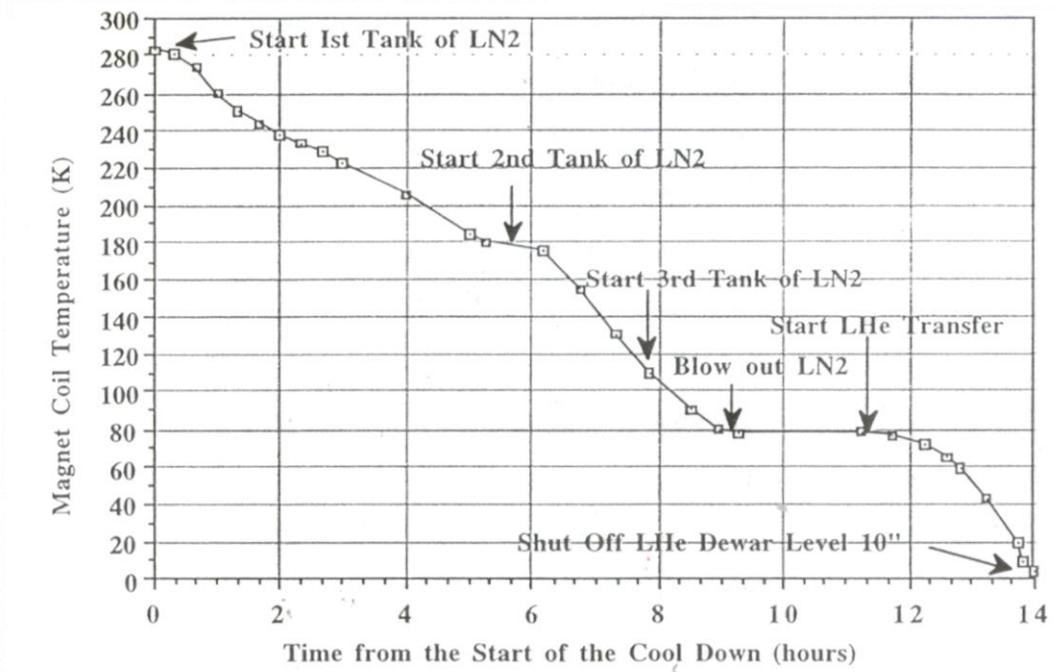

**Figure 3 Typical Cooldown with portable LN$_2$ and LHe dewars**

In 2008 Accelerator Division commissioned the MTA refrigeration system, which consisted of Sullair #1 (Brown) compressor, helium refrigerator #1 (Brown), 2-phase helium dewar, and 2-phase Mark-III helium dewar with in-line JT valve. The auxiliary system included 3,000 gal, LN2 dewar #31 and associated transfer line, 1360 ft$^3$ helium gas storage tank, inventory management system, ODH system. The following Table 1 provides summary of performance for that system.

**Table 2**

| LN2 Dewar Performance | Average boil-off rate is 60 gal/day or 2% |
|---|---|
| Sullair Compressor Capacity | Stable performance without trips/problems with average flow 45 g/s at 1.5 psig of suction pressure |
| Refrigeration Capacity | 385 Watts. Average LN$_2$ usage was 22 gal/hr (including boil-off) |
| Fill Capacity | 72 Liter/hr |

**IV Commissioning Results**

The technical description of the system, as well as the safety report, can be found in *http://mtacryo.fnal.gov*. At the time of commissioning the MTA refrigeration system consisted of Sullair #1 (Brown) compressor, helium refrigerator #1 (Brown), 2-phase helium dewar, bayonet can and transfer line interconnecting the refrigerator to the valve box installed in the hall, U-tubes and vacuum insulated flex hoses interconnecting between valve box and solenoid and cooldown equipment consisting of vaporizer, helium electric heater and piping to return cold helium back to the refrigerator suction. The auxiliary system included 3,000 gal, LN2 dewar #31 and associated transfer line, 1360 ft$^3$ helium gas storage tank and inventory management system, ODH system. The detailed P&ID drawings of the subsystems are available from *http://mtacryo.fnal.gov/*. Figures 3-5 below are the snapshots of the control system and show simplified process schematics.

**Design Highlights** (for previously commissioned equipment see reference [1])

- The transfer line carries helium and nitrogen between the refrigerator room and the detector hall at MTA. The piping system is arranged like a capital 'L' with a bayonet box at each end and an expansion box at the corner. Three vessels are connected with transfer lines. The cylindrical bayonet box in the refrigerator room provides connections to the supply dewar and refrigerator. The rectangular bayonet box in the detector hall provides valves and bayonets for connections to the magnet. The expansion box, also in the detector hall, allows thermal contraction back to the two bayonet boxes. All components are vacuum insulated and have common vacuum.
- Bayonet can (dwg. 9213.550-ME-435660) has five bayonet connections, four for helium 5K (in/return) and 40K shield (in/return), and one for liquid nitrogen. Only three connections are used: helium 5K supply and return and liquid nitrogen return.
- The transfer line (dwg. 9213.550-ME-435663) carries helium and nitrogen between the refrigerator room and the detector hall at MTA.

**Table 3**

| Description | Fluid | OD | wall | Design | Operating Temp |
|---|---|---|---|---|---|
| | | inches | inches | psig | Kelvin |
| Shield Outer | nitrogen | 4.5 | 0.12 | 60 | 77 |
| Shield Inner | nitrogen | 4 | 0.12 | 60 | 77 |
| 5K Supply | helium | 1 | 0.049 | 60 | 5 |
| 5K Return | helium | 1 | 0.049 | 60 | 5 |
| 14K Supply | helium | 1 | 0.049 | 60 | 20 |
| 20K Return | helium | 1 | 0.049 | 60 | 20 |
| Relief Valve | helium | 0.5 | 0.035 | 60 | 5 |
| Female Bayonet | Helium | 1.5 | 0.035 | 60 | 5 |

- MTA hall valve box (drawing 9213.550-ME-435320) serves to interface helium and nitrogen flows to the solenoid and return cooldown helium flow back to the refrigerator suction. Five electric cryogenic valves and one external electric valve installed with the valve box serve to re-direct helium and nitrogen flows (see schematics on Fig. 5).

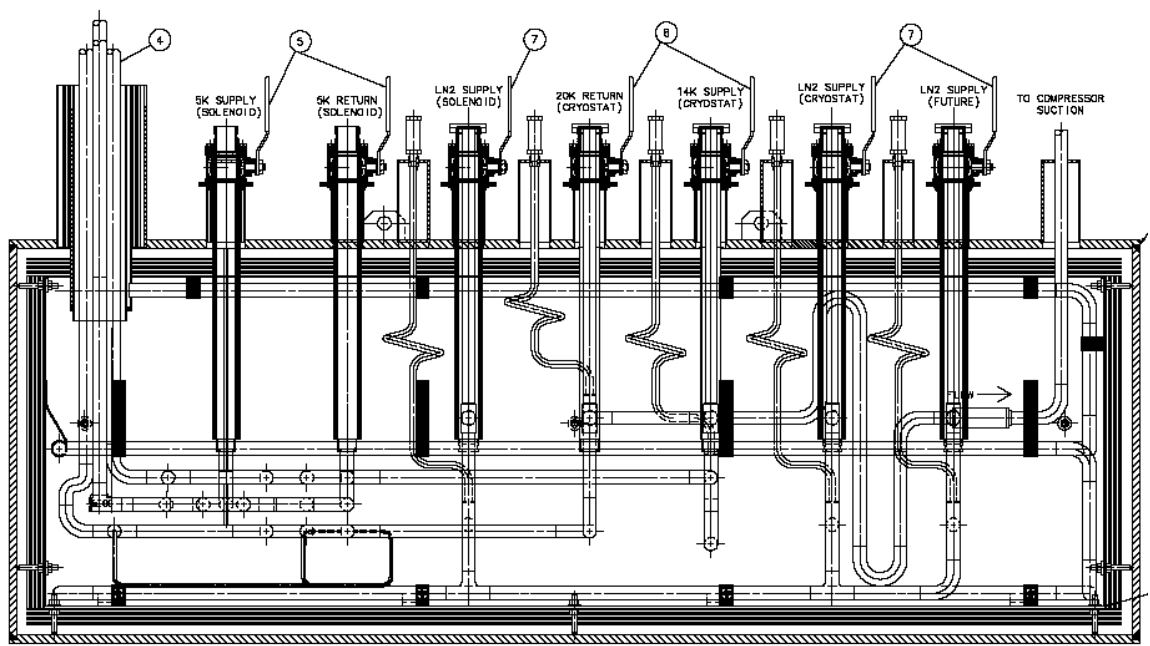

**Figure 4 MTA Hall Valve Box details**

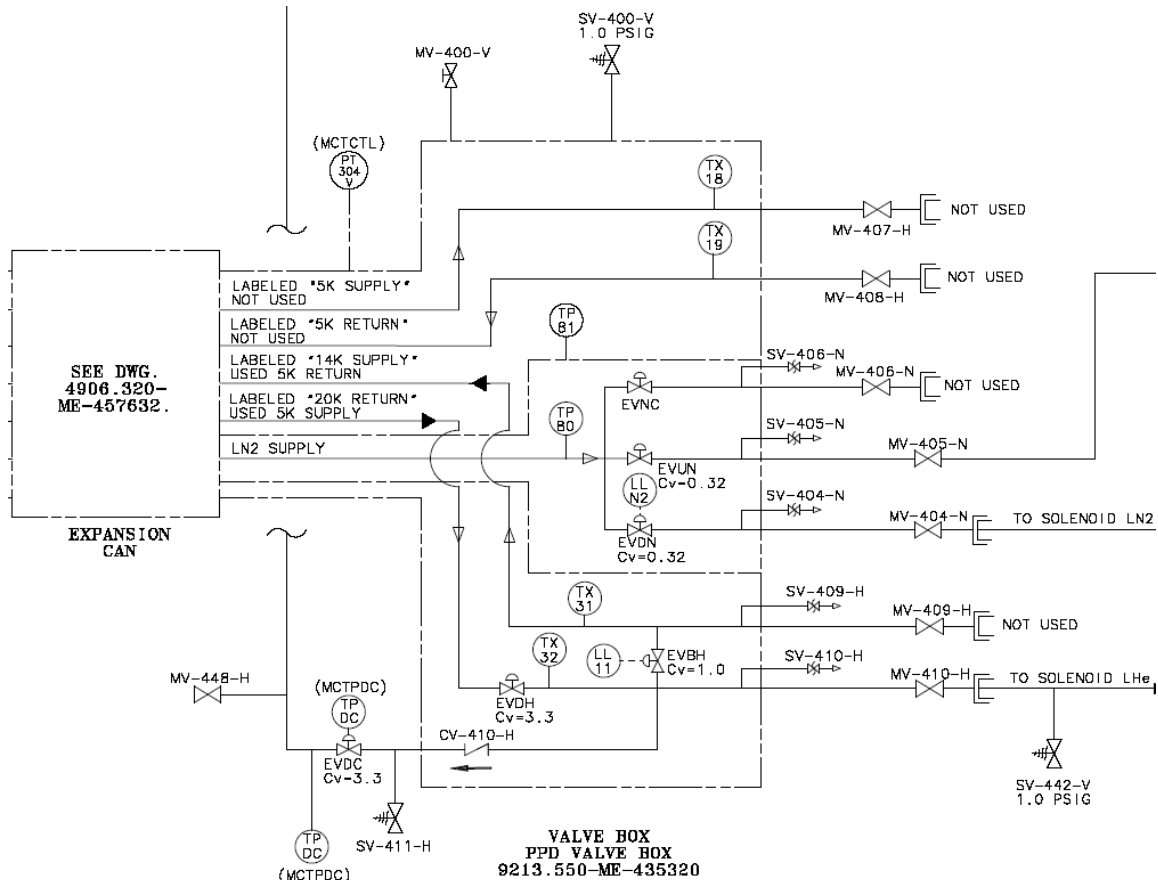
**Figure 5** MTA Hall Valve Box schematics

- In the past the solenoid was interconnected with helium (or nitrogen) supply with flexhose stinger that had to be inserted every time with detailed precision to match the helium receptacle cap on the bottom of the solenoid's helium vessel. The insertion length of the helium supply ½" stinger is 42". That stressed the bayonet connection supported with G10 material and increased probability of critical failure. The insertion length also made it impossible to un-sting helium supply after the solenoid is raised to its new specified position. In order to avoid that, a new J-type permanent stinger was built (dwg. 1650-ME-381348), installed with solid support to provide a bayonet for the helium U-tube connection from the valve box.
- New cooldown piping was designed and installed. It provided two paths for the helium return to the refrigerator suction. The first one is via bypass valve EVDC (see Fig. 5) to be used in the initial stages of the system cooldown. The second path is via a ¾" electrically actuated valve EVVT from the solenoid via vaporizer and helium heater to be used during solenoid's cooldown and operations.
- A 1 kW dual-path electric heater was installed in the experimental hall in the cooldown piping from the solenoid to suction. All electrical equipment and wiring had to be rated for operations in hydrogen environment.
- The controls for the refrigerator, including helium dewars and inventory controls for the system, are ACNET based. The schematics of the system as represented on the ACNET graphical interface are shown below on Fig. 6-8.

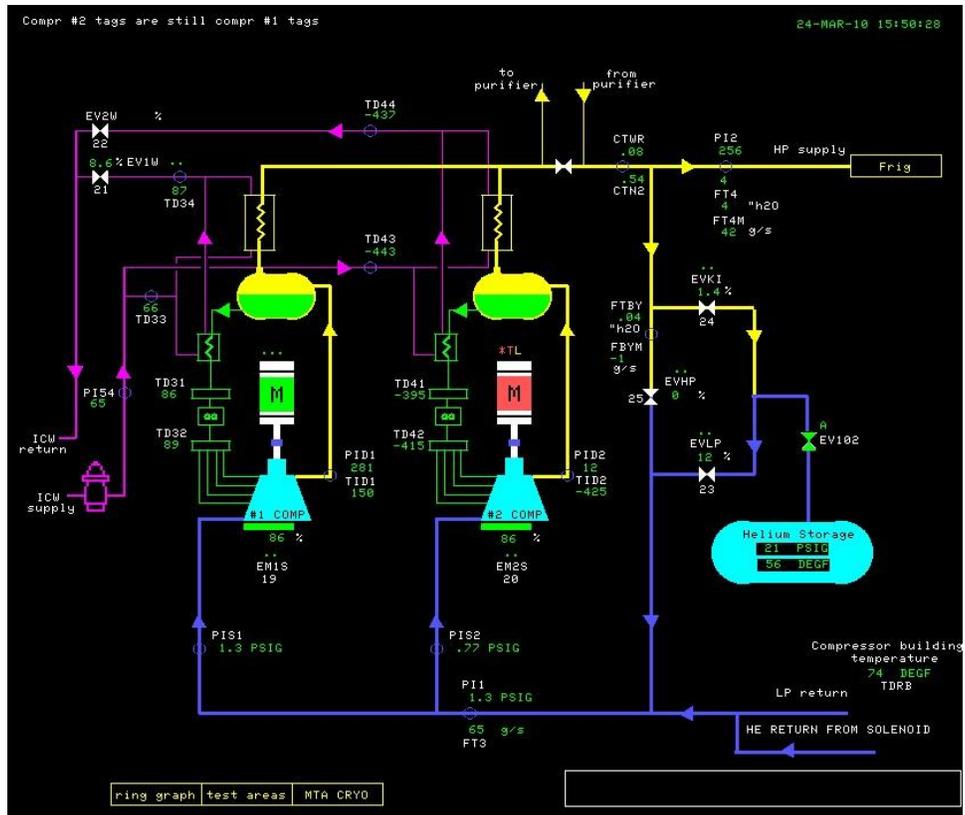

**Figure 6**

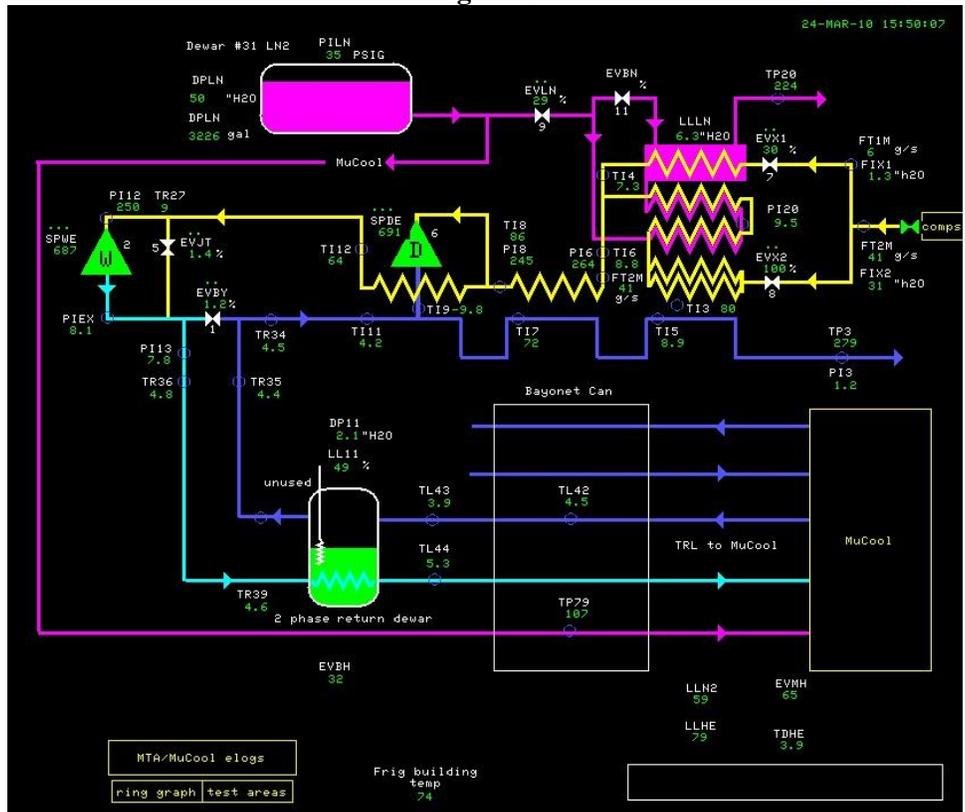

**Figure 7**

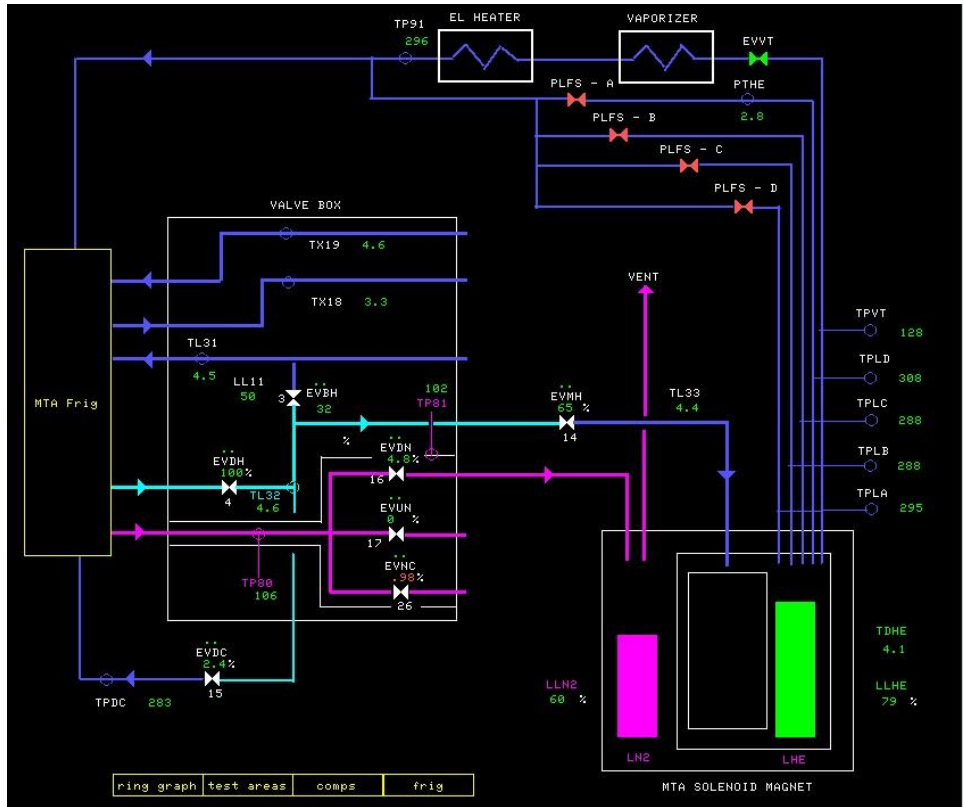
**Figure 8**

More detailed data, snap shots and data plots can be found in the MTA Cryo electronic log book http://www-mta-crl.fnal.gov/mta/Index.jsp?viewTopic=Operations/Cryo. The controls utilize I/O system and two thermometry crates typical for the Tevatron refrigerators [5]. The second thermometry crate was modified to have much lower current of 9 ua, a factor of 0.01 compare to the standard Tevatron crate in order to decrease a localized heating of the Cernox resistors, and increased the voltage gain by a factor of 100. For comparison a LakeShore module 218 was used to process the signals from Cernox resistors. A standard set of ACNET parameter pages, including F8, F9, F61, and graphical interfaces were developed to monitor the system. All parameters are data logged in 1s or 15s data loggers. The front end FrigMU CPU is located in the MTA building. Additionally, ODH chassis, Benshaw motor starter parameters and LakeShore module 218 are routed to ACNET via IRM module located in the MTA building.

**Cooldown and Fill rate**

- The cooldown of the refrigerator started from warm conditions on March 18, 2010 and took 8 hours to drop wet engine exhaust temperature below 9K.
- It took additional 4 hours, or total of 12 hours to build 60% level in the 100 liter 2-phase separator. The bypass valve EVBH was opened up to 80% to alleviate cooldown and fill. The solenoid liquid nitrogen shield level was built to 60% within first 3 hours since connecting cooldown and stayed stable through the entire operations. As the solenoid fill valve EVMH was opened during the transfer line cooldown, the fill temperature was dropping alongside the helium supply and reached 5K within first 9 hours of cooldown.
- It took additional 26 hours since reaching level in the 2-phase separator or total of 36 hours to drop the solenoid temperature (as registered by the diode TDHE) below 5K.
- It took additional 8 hours, or total of 44 hours to build 30% level LLHE in the solenoid. The solenoid temperature was ~4K and the system was fully operational.
- It took 4 hours (from 8:30 to 12:30 on March 22) to fill the solenoid from 23% to 63%. It is fairly difficult to calculate liquid volume as a function of liquid level prove linear coverage without knowing exact quality of helium, dimensions and position of the probe (as shown on Fig.1), but it maybe estimated that the fill rate was approximately as 30 liters/hour. This is half the fill rate to the on-line Mark-III dewar measured in 2008 for the same refrigerator in the liquefier mode. Certainly the higher heat losses along the different components of the system and flashing while expanding into solenoid might have contributed to lower fill rate.

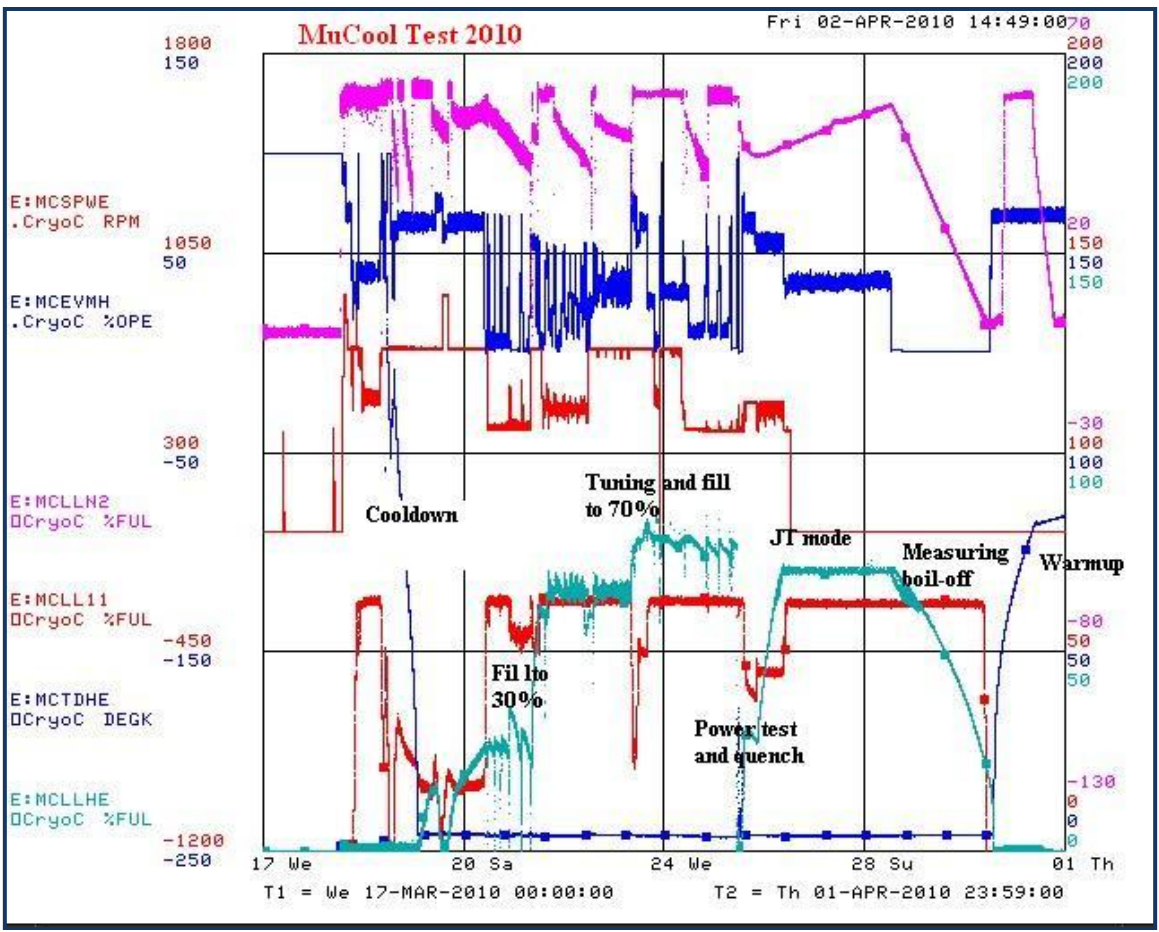

**Figure 9 MTA MuCool Refrigeration System Test Run 2010**

### Economics of Nitrogen and Power Consumption

- Assuming that the helium system is tight, then after initial cooldown and fill the main cost contributors are electricity and nitrogen.
- The system was cooled down and stably maintained with one Sullair compressor at 220 kW.
- Average liquid nitrogen boil-off rate in dewar #31 was measured in 2008 as 60 gal/day (or 2%). In 2010 the boil-off rate was measured as 40 to 50 gal/day.
- Average liquid nitrogen consumption for the refrigerator precool and solenoid shied was measured as 20 gal/hr (including boil-off). This means $LN_2$ consumption of ~450÷500 gal/day and that the 3,000 gal liquid nitrogen dewar #31 must be refilled every 4 days. For the commissioning period we selected 2-day refill period.

### Helium Inventory and Leaks

- Helium inventory required to fill and maintain the system cold at 4K is less than 400 liters or 10,640 scf of helium, thus a helium inventory tank at 150 psig should be sufficient to condense and maintain liquid helium levels.
- The 24-hr helium loss measurement was done from 14:00 on March 27 to 14:00 on March 28 while the helium levels at 2-phase dewar and solenoid were kept constant and the ambient temperature made full day-night cycle. The loss rate was calculated by the mass loss from the 1,360 $ft^3$ inventory helium tank from 23.7 psig to 16.7 psig. That loss was 0.28 lb/hr or 27 scfh. This is not a significant helium loss. Most likely it can be attributed to o-rings in the power lead flowmeters and other small leaks.

### Performance and Stability

- The system demonstrated a very stable overall performance. As shown on Fig.9, the system stably ran in JT mode without wet engine. Both dry and wet engines had to be locked in order to stay below certain maximum speed (typically below 500 rpm) in order to prevent dragging down the compressor discharge and wet engine inlet pressures. It may be beneficial to open clearances for the wet engine so to reduce mass flow through the valves. Plots on Fig.10 show clearly that helium levels in both 2-phase subcooler and solenoid could be maintained at different expander speeds, inlet or exit pressures while expanding helium in wet engine from ~5K to 2-phase.
- Since the load was maintained stably at different modes, the only optimization was to conserve liquid nitrogen. Plots on Fig. 11 show that the nitrogen usage rate is very dependent on opening of the solenoid shield valve EVDN. Most probably, the valve is oversized and it will be modified for much smaller $C_v$.
- There was only one system instability, namely sudden pressurization of the solenoid when opening solenoid helium fill valve EVMH. Plots on Fig. 12 show pressure spike PTHE up to 15 psig when opening EVMH from 20 to 50%. The loop needs tuning. An additional pressure sensor will have to be installed upstream of the EVVT solenoid valve to provide better pressure protection of the solenoid.

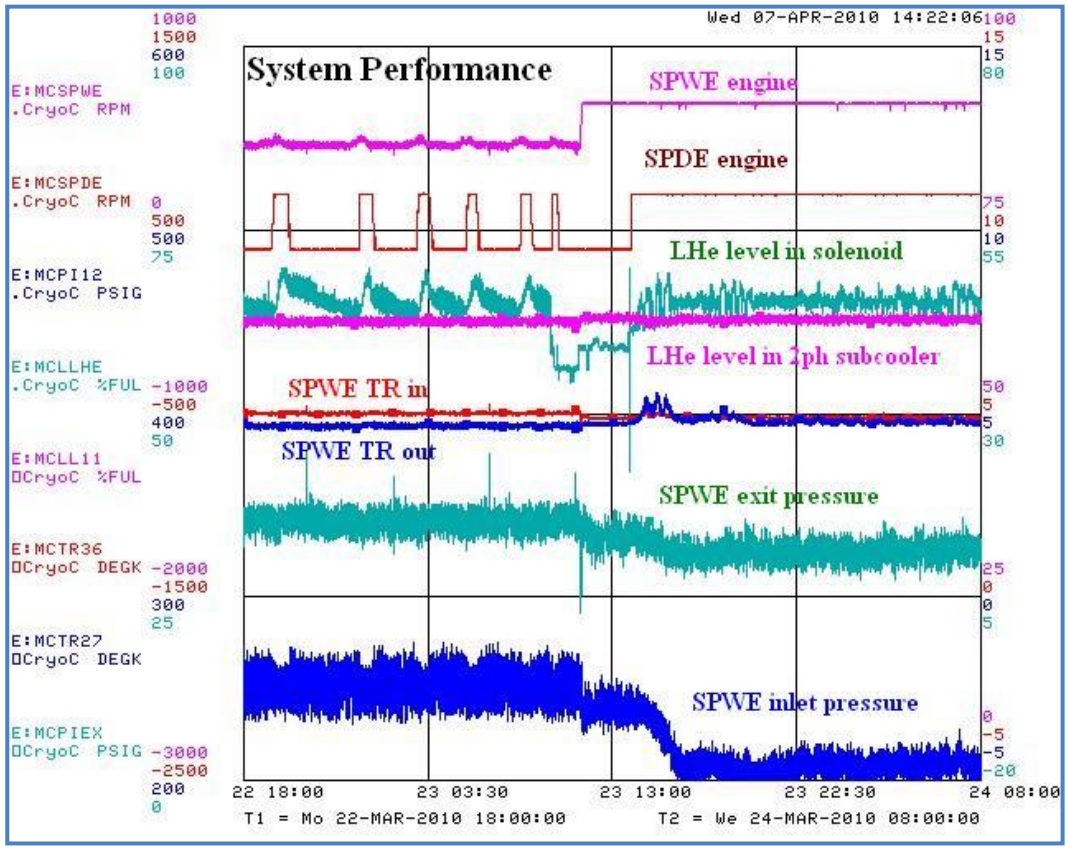

**Figure 10 System performance plots**

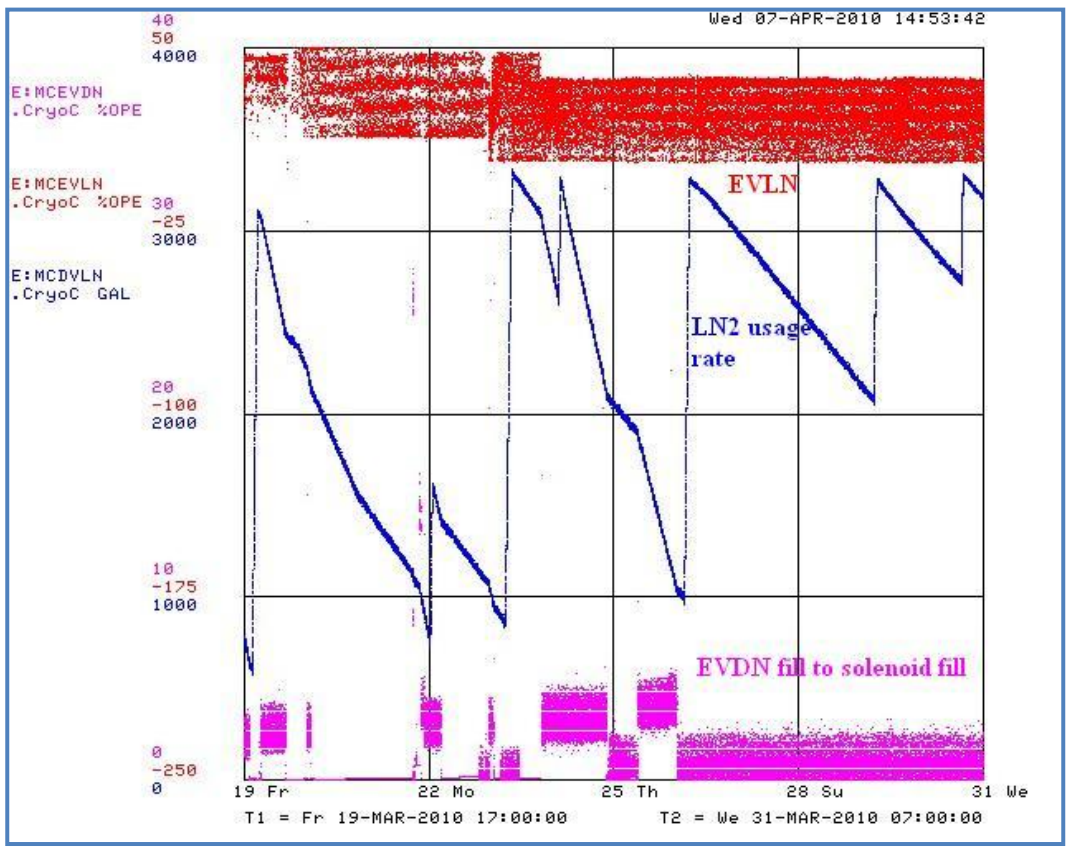

**Figure 11 Liquid nitrogen usage**

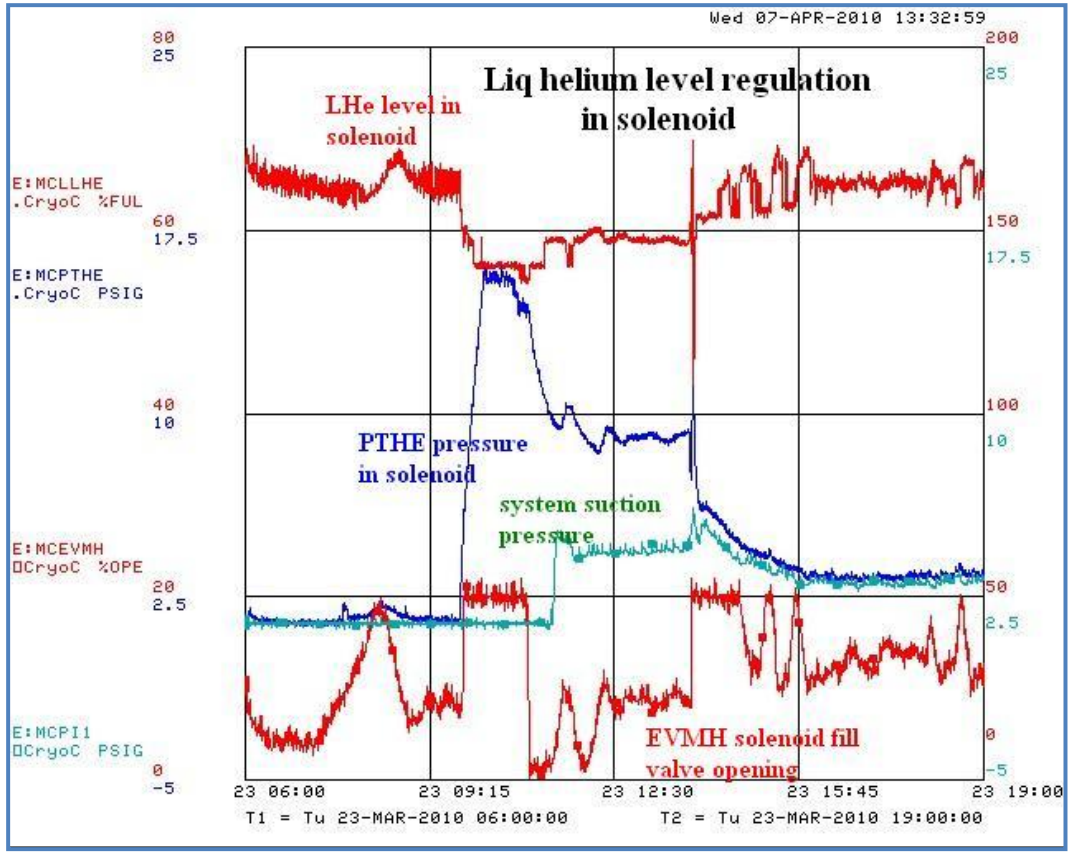
**Figure 12 Regulation of the solenoid liquid helium level**

### Quench Response

- On 03/26/10, approximately 11 am (see plots on Fig.13) the solenoid quenched at ~175 amps. That was done intentionally to verify response of the cryo system. At the time the solenoid was filled with ~200L of helium at 4K. The plots show that the solenoid pressure PTHE reached 30 psig thus rupturing the burst disk and venting through the relief vent piping to atmosphere outside the experimental hall. At the same time the ODH sensor located on the ceiling level indicated that oxygen concentration dropped to 10%. Oxygen levels were restored within 2 few minutes by hall's ventilation. All helium was supposed to be relieved to the atmosphere outside the hall. But the 3-inch relief vent line is connected to the vent line from the parallel plate vacuum relief installed on the valve box. It is done to prevent helium venting into experimental hall in that rare case if transfer line or any other component in the valve box ruptures and helium blows into vacuum space. The parallel plate relief is covered with conical reducer to catch helium and divert it to the vent line outside. During the quench event the helium flow in the vent line created pressure drop and the pressure wave back hit the parallel plate conical cover and moved it enough to relieve helium into the hall.
- After the rupture disk was replaced and the system was checked for tightness, the system was re-cooled and helium level restored without any problems.

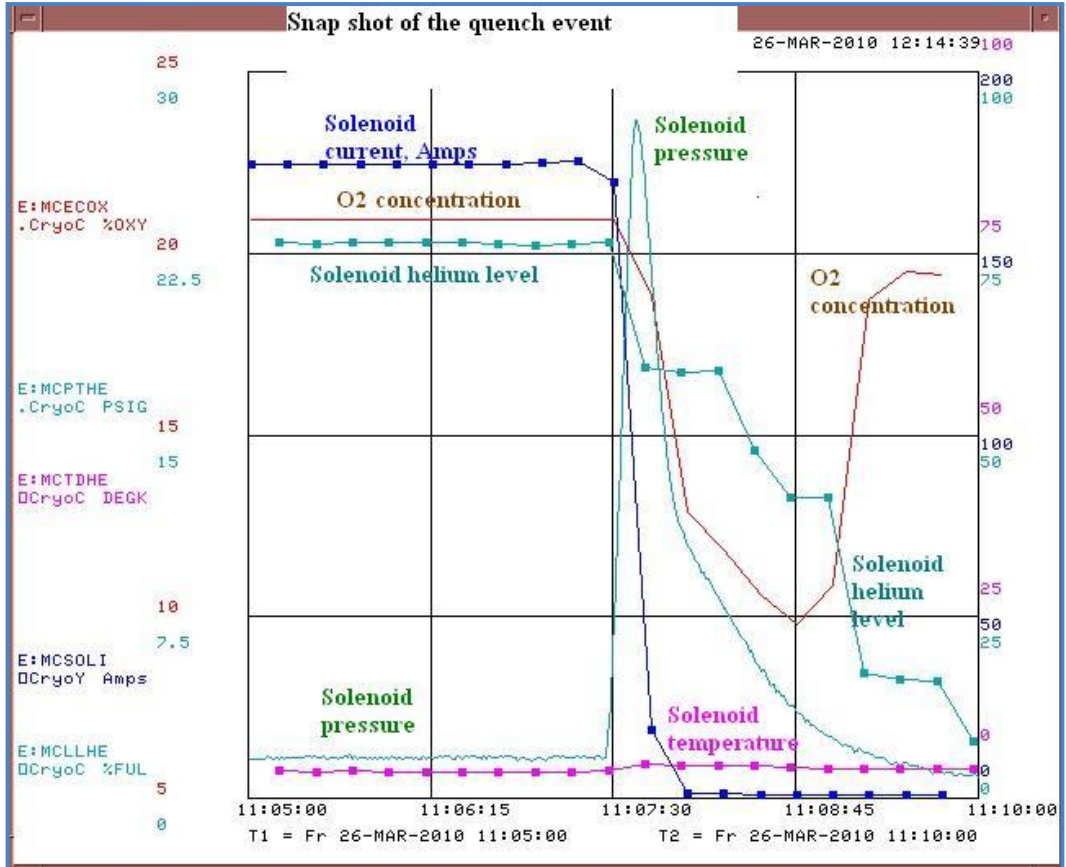
**Figure 13 Solenoid quench**

### Cernox Thermometry

Eight new Lake Shore Cernox CX-1050 resistors were installed to accurately measure temperature in the helium paths (*http://www.lakeshore.com/temp/sen/crtdts.html*). In order to measure, scale and use the temperature with Cernox resistors, we used two methods:
- Eight channel Lake Shore Model 218 monitor with two computer interfaces, IEEE-488 and serial port. The monitor required entering tabulated data in Log(Resistance, Ohm) versus Temperature (deg.K). The module then was computing the temperature in deg.K and transmitted it to ACNET via internet rack module (IRM). This solution offered precise and reliable, though relatively expensive method of using Cernox resistors.
- Tevatron-style thermometry crate modified to have much lower current of 9 ua, a factor of 0.01 compare to the standard Tevatron crate in order to decrease a localized heating of the Cernox resistors, and increased the voltage gain by a factor of 100. In order to scale measured resistance into temperature, the tabulated data for each individual Cernox resistor was fitted into equation of the following form:

$$T, degK = a1 \times 10\wedge(a2 + a3 \times Log10(R) + a4 \times Log10(R)^2 + a5 \times Log10(R)^3)$$

This fit allowed accuracy of better than +/- 2% at each temperature level over the entire range from 4K to 300K.
- The raw resistance readings from each of the eight Cernox sensors were routed though a specially designed switch box that allowed easy switching between LakeShore monitor 218 and thermometry crate. Then the readings were compared to evaluate comparative accuracy. It was found that localized heating does not produce significant error (above 0.2K at 4K level) if the switching current is reduced to 9 ua. Still, the issues of low level signal noise were persistent for the Cernox censors located in the experimental hall more than 100 feet away from the thermometry crate.

**Helium Boil-off Test**

The helium boil-off rate of the unpowered solenoid was measured on March 29-31 with fully closed both liquid helium EVMH and liquid nitrogen EVDN valves (Fig. 14). It was measured to be from 0.75 %/hr (with $LN_2$ shield above 40%) to 1.5 %/hr (with $LN_2$ shield below 40%). As it is difficult to integrate actual volume of the helium vessel over the depth of the probe, we assume the boil-off level to be 1.1 %/hr. It is important to note that the solenoid temperature stayed between 3.9K and 4.K within the whole range of LHe level. Assuming initial level of LHe in the solenoid as 70% or 200 liters, this translates to 3 Watts heat load for the unpowered solenoid with compromised nitrogen shield. By comparison, previous results indicated 3÷4 Watts boil-off for the powered solenoid with nominal helium and nitrogen levels. The solenoid can stay up to 48 hours cold and minimally filled if the nitrogen shield is maintained.

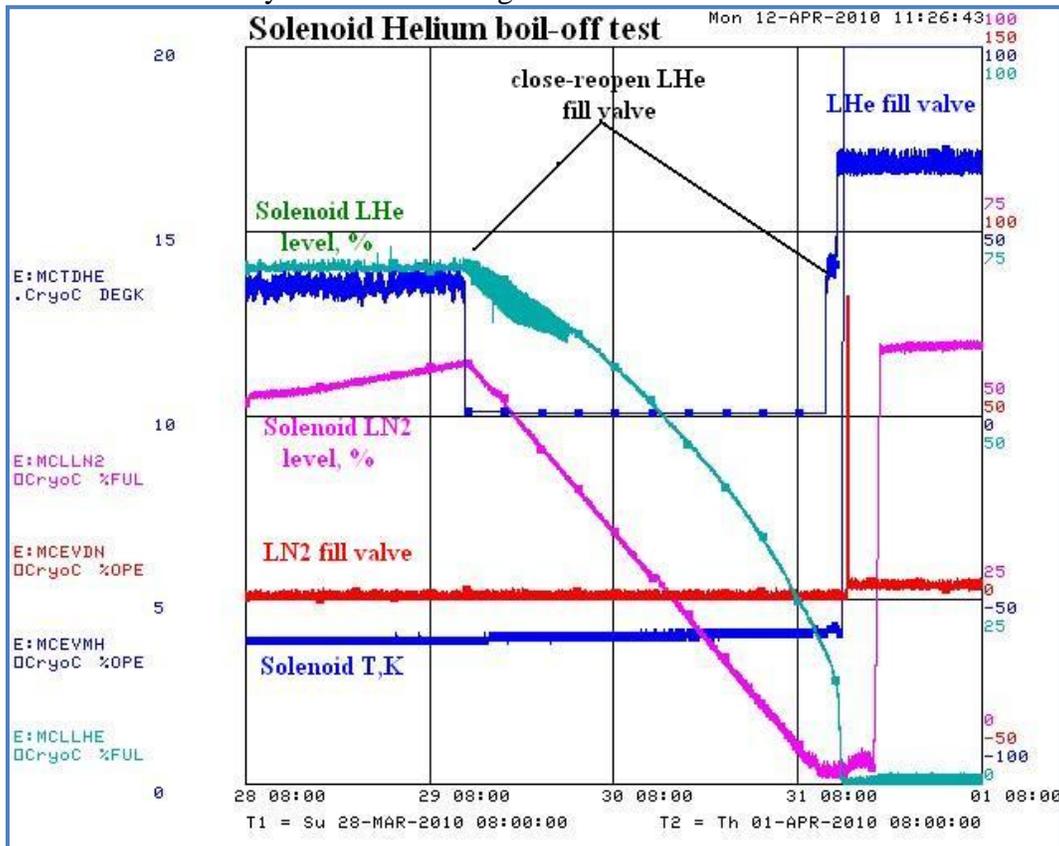

**Figure 14 Helium Boil-off Test**

## V  System Improvement List

Though the system performed well, there are several areas where additions and improvements are needed to ensure equipment safety, ease of remote operations, redundancy and accuracy. The following need to be done:
- Complete modifications and commissioning into service the second Sullair "Red" compressor. That will ensure support of the operations with much greater availability. Mechanical, electrical, controls and safety items need to be completed very similar to those done for the "Brown" compressor.
- Install a pressure transmitter in between the solenoid and the electrical cooldown valve EVVT. That pressure transducer will be able to detect pressure in the solenoid with much smaller lag. If the pressure in the solenoid spikes above 10 psig, EVVT should be opened with the additionally created finite state machine. Another finite state machine will need to be created to turn on the electric heater in the cooldown helium path.
- Replace seat and bullet of the nitrogen fill valve EVDN with smaller orifice valve and stroke it.
- Find a better tune for the helium fill valve EVMH.
- Install better thermo insulation for the exposed cold piping (nitrogen vent and helium cooldown to EVVT) as well as the top flange of the solenoid in order to minimize icing and condensation.
- Investigate electrical noise and accuracy issues for the Cernox resistors read via Tevatron-style thermometry crate.
- Create algorisms, finite state machines and graphical interface for helium inventory, startup of the compressors and system cooldown similar to Tevatron.
- Complete writing and approval of the dangerous operations procedures, operating guides and provide training to AD/Cryo/Operations for providing support for MuCool cryo operations.

## VI  Conclusion

MuCool 5T solenoid was successfully cooled down and operated coupled with MTA "Brown" refrigerator. The system performed as designed with substantial performance margin. All process alarms and interlocks, as well as ODH and fire alarms, were active and performed as designed.
- The cooldown of the refrigerator started from warm conditions and took 44 hours to accumulate liquid helium level and solenoid temperature below 5K.
- Average liquid nitrogen consumption for the refrigerator precool and solenoid shield was measured as 20 gal/hr (including boil-off).
- Helium losses were small (below 30 scfh).
- The system was stable and with sufficient margin of performance and ran stably without wet expansion engine.
- Quench response demonstrated proper operation of the relieving devices and pointed to necessity of improving tightness of the relieving manifolds.
- Boil-off test demonstrated average heat load of 3 Watts for the unpowered solenoid. The solenoid can stay up to 48 hours cold and minimally filled if the nitrogen shield is maintained.

A list of improvements includes commencing into operations the second helium compressor and completion of improvements and tune-ups for system efficiency.

## VII   Lessons Leaned

MuCool 5T solenoid was again cooled down and operated coupled with MTA "Brown" refrigerator again on May 3-7, 2010 (see Fig. 15). In the beginning, this cooldown was not successful. We attempted to cool down with wide open EVBH and EVMH fill valves for respectively 2phase dewar and the solenoid. That resulted in low supply pressure PI13 and inefficient JT-ing to the helium volumes. When this was understood, we set the EVBH to regulate PI13=17 psig and EVMH to regulate the wet engine inlet temperature TR27=9K, same way as the satellite refrigerators do.

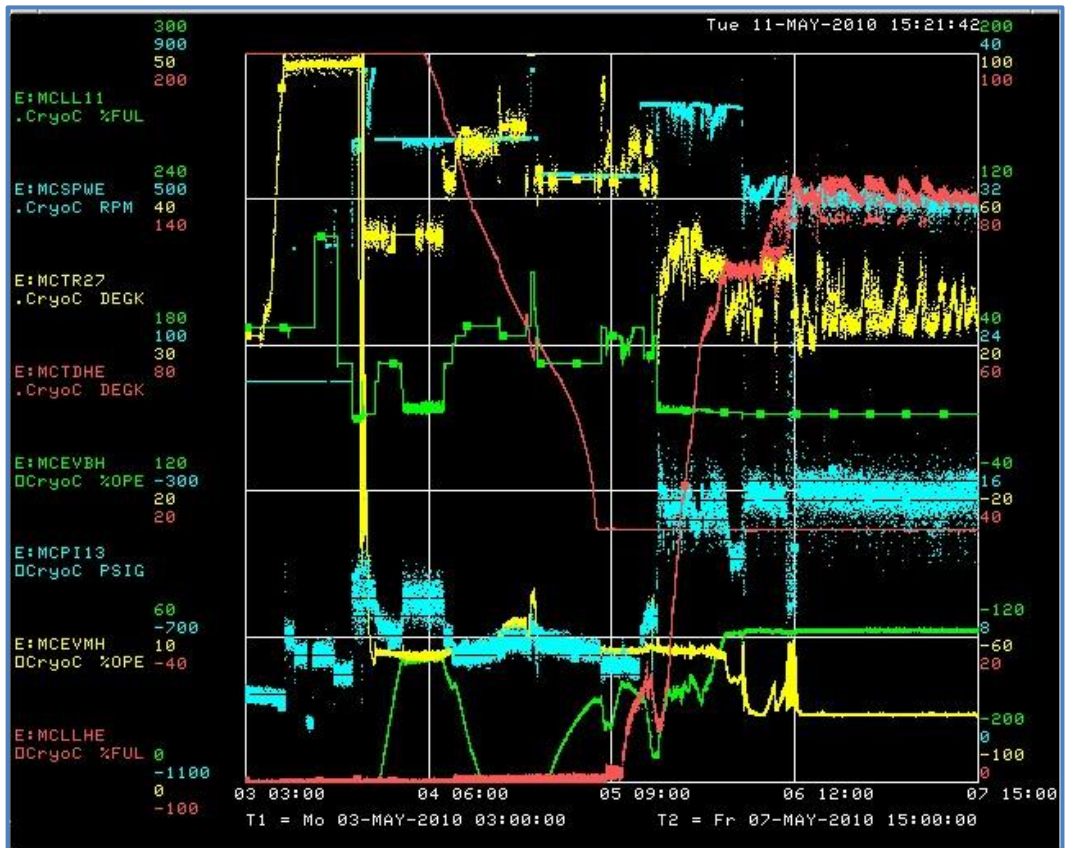

**Figure 15 Cooldown May 2010**

The system performed as designed with substantial performance margin. All process alarms and interlocks, as well as ODH and fire alarms, were active and performed as designed.
- After the system was properly configured it took 10 hrs to accumulate level to 60% MCLL11 and 80% MCLLHE.
- After the system was properly tuned up to regulate MCEVLN on MCLLLN and MCEVX1 on MCTI5 30-100%, the average liquid nitrogen consumption for the refrigerator precool and solenoid shield was measured as 15 gal/hr (including boil-off).

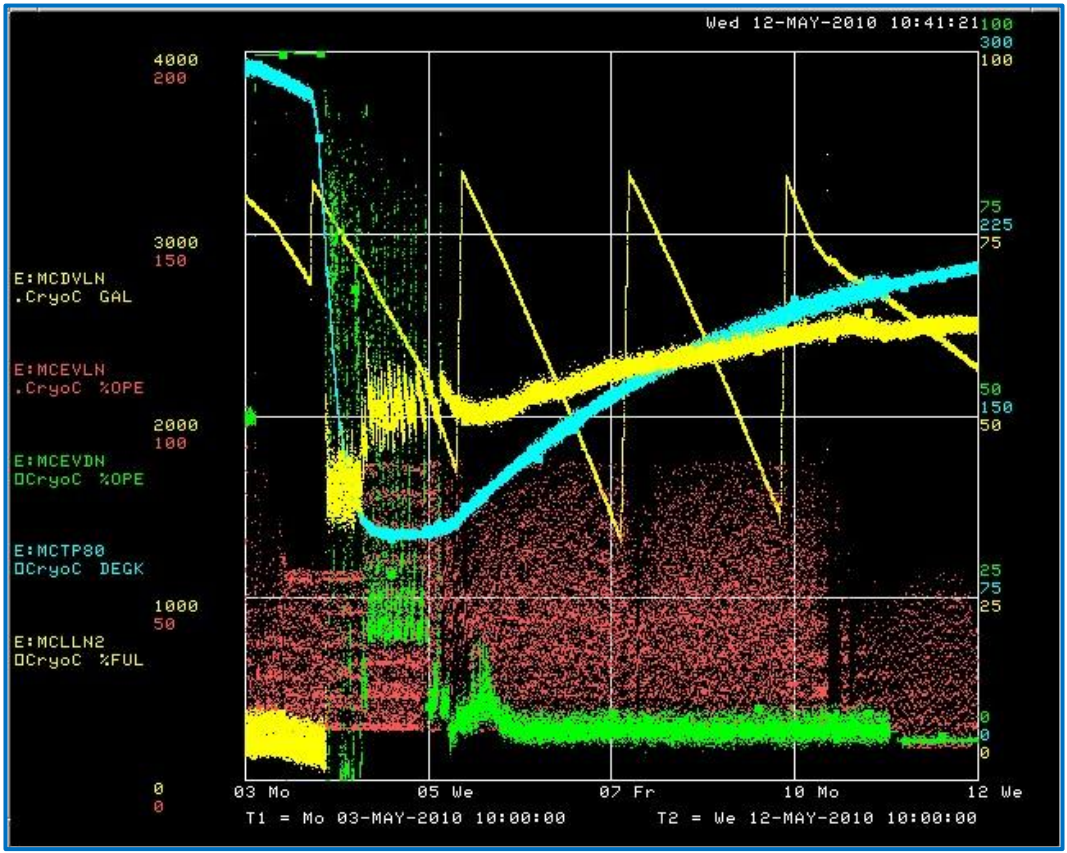

**Figure 16 LN2 consumption**

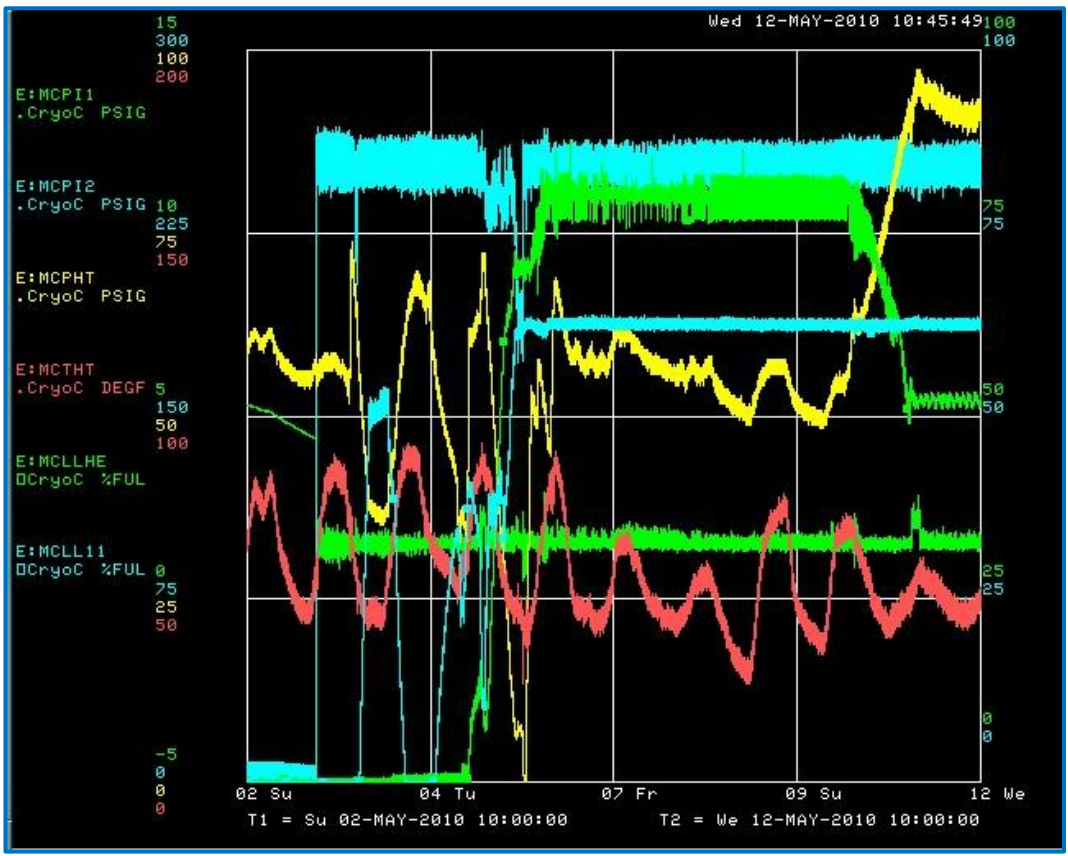

**Figure 17 Helium Loss and Levels**

Improvements proposals (05/11/2010):

After some discussions with number of people, we are proposing the following to ensure better safety and efficiency in the operations of the 5T magnet. Please review and comment. We feel that though it may result in some inconveniences or delays in operations, it will make operations easier and safer in the long run.

1. We found that cooldown solenoid EVVT must be kept open at all times to keep the solenoid pressure below 10 psig. We do not want to exercise safety reliefs to keep that pressure in check. At the same time, we also found that keeping the solenoid wide open makes the pressure differential between the solenoid and compressor suction low and insufficient to secure good flows through the power leads. We made a test this morning and saw large response in both leads flows and temperatures when we ran the solenoid pressure at 4-5 psig versus less than 2 psig at fully opened EVVT. Therefore, we propose to **a)** keep the solenoid fully opened at all times with power required for its closing and **b)** incorporate a back pressure regulator in the cooldown line back to compressor suction to keep the solenoid pressure at 4-5 psig. This is overall good solution for pressure safety; instead of throwing helium to atmosphere in case of pressure surge, we would first vent it to suction. This is an easy fix; all in the frig room, no welding.
2. We are concerned with "no-flow" conditions for the power leads. This condition can exist should the compressor suction pressure become high, e.g. if the compressor trips off. We would like to incorporate the following hardwired logic.
   - IF helium level MCLLHE is below 40% - the magnet PS is OFF
   - [[IF differential pressure (MCPTHS-MCPTHE) is below 0.5 psig] OR [IF compressor is OFF]] AND [magnet PS is ON], then the vent solenoid OPENs to blow downstream of the power leads to atmosphere. This is a medium difficulty fix in the frig room; check valve is required; no welding.
3. We found that nitrogen level in the solenoid is kept so well at helium temperature that it requires almost no supply flow. Therefore, the supply valve is kept closed. This is too bad, as this is the only path for the nitrogen from the LN2 dewar to the magnet via transfer line. With nitrogen flow OFF, the transfer line and the valve box, which are designed without shield flows, keeps warming up. Therefore, we would like to incorporate a turn-around U-tube in the hall valve box to bypass the magnet and turn around the nitrogen flow via existing available control valve back to the refrigerator via one of the non-used helium lines in the transfer line. Then that small nitrogen flow will be vented outside. The control loop will keep the shield temperature at the predefined temp level.